\documentclass[12pt]{article}
\usepackage{amsmath}
\usepackage{graphicx,psfrag,epsf}
\usepackage{enumerate}
\usepackage{natbib}
\usepackage[english]{babel}
\usepackage{epstopdf}
\usepackage{amssymb}
\usepackage{subfigure}
\usepackage{amsfonts}
\usepackage{graphicx}
\usepackage{rotating}
\usepackage[font={small},labelfont=bf]{caption}

\begin{document}

\title{\textbf{Testing for equivalence: an intersection-union permutation
solution}}
\author{R. Arboretti \\
Department of Civil, Environmental and Architectural Engineering,\\
University of Padova, Italy \\
E. Carrozzo \\
Department of Management and Engineering, University of Padova, Italy\\
F. Pesarin \\
Department of Statistical Sciences, University of Padova,Italy\\
L. Salmaso \\
Department of Management and Engineering, University of Padova, Italy}
\maketitle

$\boldsymbol{Abstract:}$ The notion of testing for equivalence of two
treatments is widely used in clinical trials, pharmaceutical experiments,
bioequivalence and quality control. It is essentially approached within the
intersection-union (IU) principle. According to this principle the null
hypothesis is stated as the set of effects lying outside a suitably
established interval and the alternative as the set of effects lying inside
that interval. The solutions provided in the literature are mostly based on
likelihood techniques, which in turn are rather difficult to handle, except
for cases lying within the regular exponential family and the invariance
principle. The main goal of present paper is to go beyond most of the
limitations of likelihood based methods, i.e. to work in a nonparametric
setting within the permutation frame. To obtain practical solutions, a new
IU permutation test is presented and discussed. A simple simulation study
for evaluating its main properties, and three application examples are also
presented.

\bigskip

\noindent \textit{Keywords:} intersection-union principle; mid-rank based
test; nonparametric combination; permutation tests

\newpage

\section{Introduction and motivation}

The idea of testing for equivalence of two treatments is widely used in
clinical trials, pharmaceutical experiments, bioequivalence and quality
control \citep[and reference therein]{r33} , \citep{r3, r1n, r2n}. In the
literature it is typically approached by the so-called \textit{%
Intersection-Union (IU) principle} \citep{r3, r4, r13, r33}. The FDA
glossary \citep{r9, r17} defines equivalence of clinical trials as: \textit{%
A trial with the primary objective of showing that the response to two or
more treatments differs by an amount which is clinically unimportant. That
is usually demonstrated by showing that the true treatment difference is
likely to lie between a lower and an upper equivalence margin of clinically
acceptable differences.}

The IU approach considers with the role of alternative hypothesis ($H_{1}$
say) that the effect of a new treatment -typically a drug- lies within a
given interval around that of the comparative treatment and with the role of
null hypothesis ($H_{0}$) that it lies outside that interval.

Without loss of generality and for the sake of simplicity, we illustrate the
proposed methodology with reference to a two-sample design and a one
dimensional endpoint variable $X\sim F,$ where the distribution $F$ is
unknown unless it is explicitly defined. Extensions to multidimensional
settings and to other designs will be the matter of further researches.
Assume that $n_{1}$ IID data are drawn from $X_{1}$ related to treatment 
\textit{A} and, independently, $n_{2}$ IID observations related to treatment 
\textit{B} are drawn from $X_{2}$. This setting can generally be obtained
when $n_{1}$ units out of $n$ are randomly assigned to \emph{A}\textit{\ }%
and $n_{2}=n-n_{1}$ to \emph{B}. We define responses as $X_{1}=X+\delta _{A}$
and $X_{2}=X+\delta _{B},$ where the underlying variable $X$ is common to
both populations where $\delta _{A}$ and $\delta _{B}$ represent the effects
of treatments \emph{A} and \emph{B,} respectively. Hence, $\mathbf{X}%
_{1}=(X_{11},\ldots ,X_{1n_{1}})$ are the data of sample \emph{A} and $%
\mathbf{X}_{2}=(X_{21},\ldots ,X_{2n_{2}})$ those of sample \emph{B}\textit{%
. }Of course, if effects are fixed, data are homoschedastic, a condition
which can considerably be weakened, see \citet{r22,r23}.

To make inference on the \textit{substantial equivalence} of two treatments,
the IU approach consists in checking if the effect $\delta _{B}$ lies in a
given interval around $\delta _{A}$. That is, by defining the difference of
effects as $\delta =\delta _{B}-\delta _{A},$ to specifically test for the
null hypothesis $H_{0}:[(\delta \leq -\varepsilon _{I})$ OR $(\delta \geq
\varepsilon _{S})]$ against the alternative $H_{1}:(-\varepsilon _{I}<\delta
<+\varepsilon _{S})$, where $\varepsilon _{I}>0$ and $\varepsilon _{S}>0$
are the non-inferior and the non-superior margins for $\delta $. Margins
that are assumed to be suitably established by biological, clinical,
pharmacological, physiological, technical or regulatory considerations. The
literature on the subject matter is quite wide and to our goal of presenting
a new permutation procedure we quote only some few relevant papers: 
\citet{r3, r4, r33, r17, r14, r20, r26, r34, r7,
r12, r27}.

Assuming that $H_{0I}:\delta \leq -\varepsilon _{I},$ $\ H_{1I}:\delta
>-\varepsilon _{I},$ $\ H_{0S}:\delta \geq \varepsilon _{S},$ and$\ \
H_{1S}:\delta <\varepsilon _{S}$ \ are the related partial sub-hypotheses,
the hypotheses of a IU test are then stated as $H_{0}=H_{0I}\bigcup H_{0S}$
against$\ H_{1}=H_{1I}\bigcap H_{1S}.$ It is worth noting that $H_{0}$ is
true if only one between $H_{0I}$ and $H_{0S}$ is true, because the two
cannot have common points; $H_{1}$ is true when both sub-alternatives $%
H_{1I}\ $and $H_{1S}$ are jointly true.\medskip\ 

In practice, the IU solution requires \emph{Two One-Sided }partial\emph{\
Tests} (TOST) \citep{r35, r36}, for instance, such as those based on
divergence of sample averages: $T_{I}=(\bar{X}_{2}+\varepsilon _{I})-\bar{X}%
_{1}$ and $T_{S}=\bar{X}_{1}-(\bar{X}_{2}-\varepsilon _{S})$, where $T_{I}$
is for testing $H_{0I}$ V.s $H_{1I}$ and $T_{S}$ for $H_{0S}$ V.s $H_{1S}$
(note that large values of both statistics are evidence against the
respective sub-null hypotheses). After then, according to the IU principle,
to obtain a global test, two partial tests must be suitably \emph{combined}
into $T_{G}=IU(T_{I},T_{S})$. A typical and effective combination is:

\begin{center}
$T_{G}=\min(T_{I},T_{S})\equiv\max(\lambda_{I},\lambda_{S})$,
\end{center}

\noindent where $(\lambda _{I},\lambda _{S})$ are two $p$-value statistics.

It is to put into evidence that two partial tests are negatively related.
Indeed, when $\varepsilon _{I}=\varepsilon _{S}=0$ they are such that $%
T_{I}+T_{S}=0$ with probability one.

\begin{figure}[tbp]
\centering
\includegraphics[scale=0.65]{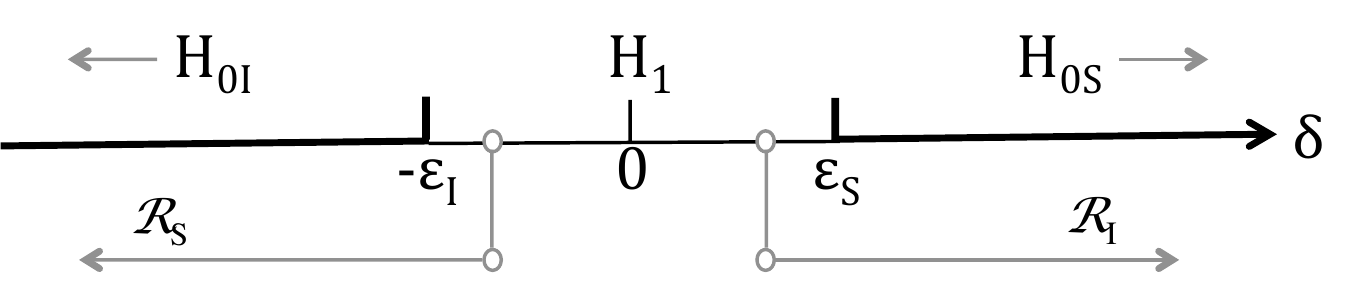}
\caption{Direction of rejection regions of IU partial tests.}
\label{Fig.1}
\end{figure}

Figure~\ref{Fig.1} presents a sketch of the IU testing situation where $%
\mathcal{R}_{I} $ and $\mathcal{R}_{S}$ represent two sub-rejection regions
in the $\delta $ axis.

\medskip 

\medskip

Of course, when either $\varepsilon_{I}$ or $\varepsilon_{S}$ is large (or
even infinitely large), measured by the $T_{G}$ distribution, the problem
becomes of \emph{non-inferiority} or \emph{non-superiority}. It is worth
noting that in such a case the testing problem becomes equivalent to a very
standard one-sided situation for a composite null against a composite
alternative, then presenting no real difficulties. Difficulties which,
instead, come out whenever $(\varepsilon_{I},\varepsilon_{S})$ and/or $%
(n_{1},n_{2})$ are not sufficiently large and that are the core problem for
testing equivalence within the IU principle.\medskip\ 

The rationale for the IU approach, which in practice is the only one adopted
in the literature for equivalence, is that \emph{equivalence is accepted if
both partial tests jointly reject}. It is, however, worth noting that this
solution mimics that connected to the well-known theorem in \citet{r15}; see
also: \citet{r33, r26}. This essentially states that, \emph{under very
stringent assumptions one unconditionally likelihood-based optimal}(UMPUI) 
\emph{test\ }$T_{Opt}$\emph{\ exists}.

Denoting by $\phi _{T}$ and $\phi _{h}$ the indicator functions of rejection
regions of tests $T$ and $T_{h},h=I,S,$ respectively, with clear meaning of
the symbols such a solution is optimal within the class of tests $\mathcal{T}
$ that satisfy the conditions: a) $\sup_{\delta \in H_{0}}[\mathbf{E}%
_{F}(\phi _{T},\delta )]\leq \alpha ,$ i.e. $T$ is at most of size $\alpha ;$
and b) $\inf_{\delta \in H_{1}}[\mathbf{E}_{F}(\phi _{T},\delta )]\geq
\alpha ,$ i.e. $T$ is at least unbiased. So, it is required that every such
global test $T\in \mathcal{T}$ has type I error rate \emph{not larger than }$%
\alpha $. And thus each $T$ has to satisfy $\alpha $ at both extremes of $%
H_{1}.$ That is: $\mathbf{E}_{F}(\phi _{T},\varepsilon _{h})\leq \alpha $ at 
$\varepsilon _{h}=-\varepsilon _{I},\varepsilon _{S},$ and $\mathbf{E}%
_{F}(\phi _{T},\delta )\geq \alpha ,$ at $-\varepsilon _{I}<\delta
<+\varepsilon _{S}.$ As a consequence, each partial test $T_{h},$ $h=I,S,$
must be \emph{calibrated} \citep{r26} so as their IU combination must
satisfy both conditions. This leads to define calibration, expressed in
terms of partial type I error rate $\alpha ^{c}$, by means of the equation:

\begin{center}
\begin{tabular}{c}
$\alpha^{c}=\mathbf{E}_{F}(\phi_{h},\varepsilon_{h})~,~~$%
\end{tabular}
\end{center}

\noindent under the condition

\begin{center}
\begin{tabular}{c}
$\mathbf{E}_{F}(\phi_{T},\varepsilon_{h})=\alpha,~h=I,S.$%
\end{tabular}
\end{center}

It is worth noting that calibrated $\alpha ^{c}$ is common to both margins,
because it essentially depends on the equivalence interval length $%
\varepsilon _{I}+\varepsilon _{S}$, and on the distributions associated to
the specific partial tests $T_{h}$ and global test $T$ under consideration.
More specifically, it depends on $F$, on partial tests $T_{h}$ and on global 
$T$ through their rejection regions $\phi _{h}$ , $h=I,S,$ and $\phi _{T}$,
where this latter to be defined requires the knowledge of the solution $%
\alpha ^{c}$. Thus, the calibration is generally not a simple process.
Indeed, a rather intriguing mathematical problem comes out, since we can say
that \emph{to obtain calibrated }$\alpha ^{c}$\emph{, in practice one has to
know it}. According to \citet{r15}, this calibration can be achieved via
numeric calculations if the underlying distribution $F$ lies within a
uniparametric or even a bi-parametric regular exponential family, if for the
latter the invariance property works for one nuisance parameter \citep{r33}.
In other cases, it has to be obtained via Monte Carlo simulations under the
conditions stated at the following point ii) because, to the best of our
knowledge, direct numeric calculations are not available.

In practice, the IU $T_{G}$ rejects at global type I error rate $\alpha $ if 
$\max (\lambda _{I},\lambda _{S})\leq \alpha ^{c}$. Such a condition is not
always simple to fulfill because, in order to establish if with actual data $%
\mathbf{X}$ both partial tests $T_{I} $ and $T_{S}$ do reject, it entails to
know the distribution function of global $T_{G}$, that depends on the
underlying $F$, on two margins $(\varepsilon _{I},\varepsilon _{S}),$ on
sample sizes $(n_{1},n_{2}),$ and two statistics $(T_{I},T_{S})$. The
central difficulty for finding the IU $T_{G}$ distribution is that two
partial tests $T_{I}$ and $T_{S}$ are negatively dependent and their
dependence, which in turn depends on the $T_{G}$ measure of $\varepsilon
_{I}+\varepsilon _{S}$, is generally much more complex than linear. This
becomes quite compelling for multivariate settings where regressions are
generally more complex than pairwise linear and so it is practically
impossible to properly manage estimators of all related coefficients, the
number and type of which are essentially unknown. 
In the literature such issues have been pointed out by \citet{r30} and \citet{r5n}. 
In these conditions, a general solution could be found if we were able to 
nonparametrically manage that underlying dependence. This is possible if we
stay within the permutation testing principle and more specifically within 
the \emph{NonParametric Combination\ }(NPC)\emph{\ of dependent Permutation 
Tests} (PTs) \citep{r37, r38, r21}.

The permutation testing principle essentially requires that in the space of
effects $\delta $ there is a point\emph{\ }$\delta _{0}\notin H_{1}$ such
that data permutations are equally likely (generally, but not always, this
corresponds to the data exchangeability property). In particular, PTs and
the NPC take benefits from the \emph{conditional and unconditional uniform
monotonicity} property. Roughly speaking, this can be referred to as: \emph{%
testing for} $H_{0}^{\dag }:\delta \leq \delta _{0}$ V.s $H_{1}^{\dag
}:\delta >\delta _{0}$ \emph{by any unbiased PT }$T,$\emph{\ with rejection
region indicator }$\phi _{T},$ \emph{such a property states that for any} $%
\delta ^{\prime }<\delta _{0}<\delta <\delta ^{\prime \prime },\ $\emph{any
data }$\mathbf{X,}$ \emph{any sample sizes }$(n_{1},n_{2})\geq 2$,\emph{and
any underlying distribution }$F,$ \emph{the following relations respectively
hold:}

\begin{center}
$%
\begin{tabular}{c}
$\lambda_{T}(\mathbf{X(}\delta^{\prime}\mathbf{)})\overset{d}{\geq}\lambda
_{T}(\mathbf{X(}\delta_{0}\mathbf{)})\overset{d}{\geq}\lambda_{T}(\mathbf{X(}%
\delta\mathbf{)})\overset{d}{\geq}\lambda_{T}(\mathbf{X(}\delta^{\prime%
\prime}\mathbf{)})$\qquad%
\end{tabular}
$
\end{center}

\noindent \emph{and}

\begin{center}
\begin{tabular}{c}
$\mathbf{E}_{F}(\phi_{T},\delta^{\prime})\leq\mathbf{E}_{F}(\phi_{T},%
\delta_{0})=\alpha\leq\mathbf{E}_{F}(\phi_{T},\delta)\leq\mathbf{E}%
_{F}(\phi_{T},\delta^{\prime\prime})$~,%
\end{tabular}
\end{center}

\noindent where: $\lambda _{T}(\mathbf{X(}\cdot \mathbf{)})=\Pr \{T[\mathbf{X%
}^{\ast }(\cdot )]\geq T[\mathbf{X}(\cdot )]|\mathbf{X}(\cdot )\}$ represent
permutation $p$-value statistics of test statistic $T$ on data sets $\mathbf{%
X(}\cdot \mathbf{)}$ with effect $(\cdot )\mathbf{,}$ $\mathbf{X}^{\ast
}(\cdot )$ being a random permutation of $\mathbf{X}(\cdot );$ moreover, due
to discreteness of permutation distributions, the $\alpha $-values are those
that are really attainable.\medskip\ 

The IU-TOST approach, as well as the likelihood-based one, presents some
serious pitfalls, as we will see while analyzing simulation results of our
permutation approach (Section 3). Most important are:

i) It does not admit any solution when $\ \varepsilon _{I}=\varepsilon
_{S}=0,$ that is when the null hypotheses is $H_{0}:[(\delta \leq 0)\bigcup
(\delta \geq 0)]$ in which case the alternative $H_{1}$ becomes logically
impossible since it is empty, $H_{1}=\varnothing $ say.

ii) Unless the invariance property works, to obtain via Monte Carlo
simulations the IU-TOST\emph{\ }$T_{G}\ $calibrated, in practice it is
required the complete knowledge of underlying distribution $F$ of data $X$,
including all its nuisance parameters. When, for partial test distributions,
a central limit theorem is working, calibrated $\alpha ^{c}$ can be
approximately determined according to \citet{r33}, since the interval length 
$\varepsilon_{I}+\varepsilon_{S}$ can be measured in terms of underlying
standard error $\sigma_{X}[n_{1}n_{2}/(n_{1}+n_{2})]^{1/2}.$

iii) When the $T_{G}$ measure of $\varepsilon_{I}+\varepsilon_{S}$ is small
there still remain severe difficulties to establish equivalence when it is
true.

iv) According to \citet{r11}, we will see that our IU permutation test $%
T_{G}=\min (T_{I},T_{S})$ quickly converges to $T_{Opt}$ in the conditions
for the latter.

v) Unless $\min (n_{1},n_{2})$ or$\ \varepsilon _{I}+\varepsilon _{S}$ are
very large, once the equivalence is rejected, the application of multiple
testing techniques for establishing which $H_{0h}$ is active, if not
impossible, is generally \emph{difficult} since calibrated $\alpha ^{c}$ lie
in the half-open interval $[\alpha ,~(1+\alpha )/2).$

vi) While using ranks, only within our permutation approach it seems
possible to express margins in terms of the same physical unit of
measurement of the data $X$ \citep{r1, r39}. Indeed, expressing them in
terms of rank transformations implies considering something similar to
random margins, the meaning of which become doubtful or at least
questionable.\medskip\ 

The IU-TOST solution usually considered in the literature \citep{r4}
corresponds to the non-calibrated version $\ddot{T}_{G}$, that which rejects
global $H_{0}$ at type I error rate $\alpha $ when both partial tests reject
each at the same rate $\alpha $ in place of calibrated $\alpha ^{c},$ i.e.
when $\ddot{\alpha}_{I}=\ddot{\alpha}_{S}=\alpha .$ This heuristic and naive 
$\ddot{T}_{G}$ solution has several further specific pitfalls:

I) It satisfies Lehmann's condition a) but not b); by the way, it trivially
satisfies Theorem 1 of \citet{r3}.

II) When the $T_{G}$ measure of $\varepsilon_{I}+\varepsilon_{S}$ is very
large, the non-calibrated naive $\ddot{T}_{G},$ whose partial type I errors
are $\ddot{\alpha}_{I}=\ddot{\alpha}_{S}=\alpha,$ and the calibrated $T_{G}$
coincide, and so they both are consistent (Section 2.4). Indeed, if $T_{I}$
and $T_{S}$ are consistent partial tests \citep{r24} and the central limit
theorem is approximately working, as sample sizes increase the $T_{G}$
measure of $\varepsilon_{I}+\varepsilon_{S}$ increases being it measured in
terms of $\sigma_{X}[n_{1}n_{2}/(n_{1}+n_{2})]^{1/2}$.

III) The naive TOST$\ \ddot{T}_{G}$ can be dramatically conservative since
its maximal rejection probability can be much smaller than $\alpha ,$ even
close to zero, as we will see.

IV) Theorem 2 in \citet{r3}, essentially states that there exist margins $%
(\varepsilon _{I},\varepsilon _{S})$ such that the power of naive $\ddot{T}%
_{G}$ is not smaller than $\alpha .$ That, however, is not a constructive
condition and so is not beneficial for finding practical solutions. Indeed,
in any real problem, based on technical or biological or regulatory
consideration, margins are established prior to the experiment for
collecting data is conducted, and not after data are collected, with the aim
of conferring the unbiasedness property to the naive TOST$\ \ddot{T}_{G}$.

V) Paradoxically, when interval length $\varepsilon _{I}+\varepsilon _{S}$
is small in terms of $T_{G}$ distribution, \emph{the maximal probability}
for the naive TOST\emph{\ }$\ddot{T}_{G}$\emph{\ to find a drug equivalent
to itself can be about zero}. This is especially true when both partial
rejection regions are external to the equivalence interval defined by $%
H_{1}, $ i.e. when $(-\varepsilon _{I},\varepsilon _{S})\bigcap [\mathcal{R}%
_{S}\bigcup \mathcal{R}_{I}]=\emptyset $.

VI) As a consequence, $\ddot{T}_{G}$ is not a member of class $\mathcal{T},$
and so in our opinion there are no rational reasons for taking it into
consideration for testing equivalence.\medskip\ 

It is worth noting that the FDA definition of testing for equivalence is
compatible also with a sort of \emph{dual formulation} (mirror-like) to that
commonly considered in the literature \citep{r25}. Indeed, the roles of null
and alternative hypotheses can be reversed. As a matter of facts, we could
rationally also consider $\tilde{H}_{0}:(-\varepsilon _{I}\leq \delta \leq
+\varepsilon _{S})$ against the alternative $\ \tilde{H}_{1}:[(\delta
<-\varepsilon _{I})\bigcup (\delta >\varepsilon _{S})].$ We are not
interested, here, to provide a comparison of two formulations, essentially
because we have to firstly discuss the permutation solution to the standard
formulation and to examine some of its performances. Such a comparison, or
better such a parallel analysis, will be the subject matter of a further
specific research.

The remainder of this paper is organized as follows: Section 2 is entirely
devoted to develop our IU-TOST-NPC method; Section 3 contains a simple
simulation study with the aim of assessing performances and pitfalls of the
IU approach; with the role of putting into evidence that IU-NPC requires
large margins to detect equivalence, Section 4 contains the discussion of
three examples: one in which two-sample data are essentially equivalent in
distribution, one near to practical equivalence, and one in which a
non-equivalence is empirically evident; finally, some concluding remarks are
in Section 5.

\section{The nonparametric IU permutation test}

For testing $H_{0}$ against $H_{1}$ within the IU our proposal is to test
separately, but simultaneously, $H_{0I}$ against $H_{1I}$ and $H_{0S}$
against $H_{1S}.$ For the sake of generality, let us suppose that data $Y$
are really observed and that margins $(\varepsilon_{I},\varepsilon_{S})$ are
expressed in the same physical units of measurements of the data. So, $(%
\mathbf{Y}_{1},\mathbf{Y}_{2})$ are two observed data sets. For testing $%
H_{0I}$ against $H_{1I}$ and $H_{0S}$ against $H_{1S},$ let us consider the
data transformations $\mathbf{Y}_{I1}=\mathbf{Y}_{S1}=\mathbf{Y}_{1},$ $%
\mathbf{Y}_{I2}=\mathbf{Y}_{2}+\varepsilon_{I},$ and $\mathbf{Y}_{S2}=%
\mathbf{Y}_{2}-\varepsilon_{S}.$ Thus, one unidimensional observed variable $%
Y$ is transformed into a two-dimensional one $(Y_{I},Y_{S}),$ where two
components are deterministically related.

A basic assumption for conferring the conditionally and unconditionally
unbiasedness to permutation partial tests \citep{r22} is that the underlying
variable $Y$ is provided with the so-called dominance in distribution
property\ with respect to the effect $\delta $. This implies that for every
real $t$ two cumulative distribution functions are related either as $%
F_{Y_{2}}(t)\leq F_{Y_{1}}(t)$ or as $F_{Y_{2}}(t)\geq F_{Y_{1}}(t)$, the
equality $\forall t$ being satisfied only in one point $\delta _{0}\notin
H_{1}.$ According to this, we have to assume that two cumulative
distributions do not intersect. Of course, this is trivially satisfied when
treatment effects are fixed. It can also be satisfied for most random effect
models, in which case for $\delta \neq \delta _{0}$ there might be
non-homoschedasticities in the data. It may be not satisfied in some
problems where treatment effects can interact with some underlying genetic
configuration \citep{r5}. It is worth noting, however, that in $\delta
=\delta _{0},$ $\mathbf{Y}_{1}$ and $\mathbf{Y}_{2}$ being equal in
distribution, data are exchangeable. Within the permutation theory random
effects are only required to be either non-negative or non-positive with
probability one, without requiring for them the existence of moments of any
order.

Suppose now that partial tests are based on divergence of sample means of
suitable transformations of the data, such as: $X=\Psi(Y),$ where $\Psi
=[\log(Y),$ $\sqrt{Y},$ $Rank(Y)$, $AUC,$ the \emph{identity }$Y$, etc.].
Thus, two partial tests assume the general form: $T_{I}=\bar{X}_{I2}-\bar {X}%
_{I1}$ and $T_{S}=\bar{X}_{S1}-\bar{X}_{S2},$ where $\bar{X}%
_{hj}=\sum_{i\leq n_{j}}X_{hji}/n_{j},$ $j=1,2,$ $h=I,S,$ are sample means.
This is of particular interest when working with rank or log transformations %
\citep{r1}.

For the sake of simplicity and without loss of generality, let us refer to
the \emph{identity} transformation, i.e. $X=Y.$ The permutation test $T_{I},$
for $H_{0I}$ against $H_{1I}$, is based on comparison of two sample means,
where the data $\mathbf{X}_{2}$ of sample \textit{B} are modified to $%
\mathbf{X}_{I2}=\mathbf{X}_{2}+\varepsilon _{I}$, while those of sample 
\textit{A} are retained as they are, i.e. $\mathbf{X}_{I1}=\mathbf{X}_{1}.$
In this way, we may write $H_{1I}:\delta >-\varepsilon _{I}\equiv X_{I2}%
\overset{d}{>}X_{I1}$ and $H_{0I}:\delta \leq -\varepsilon _{I},$ where $%
X_{I2}\overset{d}{>}X_{I1}$ emphasizes that\ $X_{I2}$ under $H_{1I}$ is
larger in distribution than $X_{I1},$ i.e.$F_{X_{2}}(t)\leq F_{X_{1}}(t)$;
instead in $\delta =-\varepsilon _{I},$ being $F_{X_{2}}(t)=F_{X_{1}}(t),$ $%
\forall t,$ data are exchangeable. Thus, the Rejection Probability\ (RP) of $%
T_{I}$ is $\alpha $ at $\delta =-\varepsilon _{I}.$ Since $\delta <\delta
^{\prime }$ implies$\ \mathbf{E}_{F}(\phi _{I},\delta )\leq \mathbf{E}%
_{F}(\phi _{I},\delta ^{\prime })$ [i.e. RP is conditionally and
unconditionally monotonic in $\delta $ ], RP is not larger\ than $\alpha $
at $\delta <-\varepsilon _{I}$, and not smaller than $\alpha $ at $\delta
>-\varepsilon _{I}$. And this uniformly for all sample data $\mathbf{X}$ \
and all underlying distributions $F$. Correspondingly, for testing $%
H_{0S}:\delta \geq \varepsilon _{S}$ against $H_{1S}:\delta <\varepsilon
_{S}\equiv X_{S2}\overset{d}{<}X_{S1}$ we use the test statistic $T_{S}=\bar{%
X}_{S1}-\bar{X}_{S2},$ where $\mathbf{X}_{S1}=\mathbf{X}_{1}$ and $\mathbf{X}%
_{S2}=\mathbf{X}_{2}-\varepsilon _{S}.$

It is worth observing that \textit{large values of both partial test
statistics }$T_{I}$ \textit{and }$T_{S}$\textit{\ are significant}. So, two
partial tests lead to $p$-value like statistics $\lambda_{I}$ and $\lambda
_{S}$ (as defined at step 8 of the algorithm \ref{Algorithm}) that are
smaller in distribution under $H_{1}$ than under $H_{0}$. It is also to be
observed that as $H_{0I}$ true implies $H_{0S}$ false, and vice versa, i.e.
two null sub-hypotheses cannot jointly be true; whereas two sub-alternatives 
$H_{1I}$ and $H_{1S}$ can. This fact implies that two partial $p$-values
statistics are negatively dependent. Such a property has to be accurately
taken into consideration while defining the $T_{G}$ distribution and while
discussing its properties.

It is important to note that since for positive variables, $X\overset{P}{>}0$
say, two test statistics $\bar{X}_{h1}-\bar{X}_{h2}\ $and $\bar{X}_{h1}/\bar{%
X}_{h2},$ $h=I,S,$ are permutationally equivalent \citep{r22}, then \emph{%
for such variables difference intervals and ratio intervals have the same
handling within the permutation setting}. There is, however, a difference in
the physical meaning assigned to margins: a) in testing by difference of
sample means, margins are expressed in the same physical units of
measurements of data $X$ and their length is evaluated, at least
approximately, in terms of standard deviation $\sigma _{X};$ b) in testing
by ratio of means, they are expressed in terms of percent of mean $\mu _{X}.$
Thus, it is not always possible to find a meaningful correspondence between\
two meanings. In any case, it is to put into evidence that the equivalence
interval length is always measured in terms of the $T_{G}$ distribution.

Following the spirit of the NPC methodologies, once two partial tests and
related $p$-value statistics $(\lambda _{I},\lambda _{S})$ are obtained, 
\textit{we must suitably combine them }so as to infer which in the light of
the data $\mathbf{X}$ between $H_{0}$ and $H_{1}$ is to be retained with
type I error rate not exceeding a given $\alpha $-value. This combination
can be done by a nonparametric combining function $\varphi
:[0,1]^{2}\rightarrow \mathbb{R}^{+},$ \textit{small values of which are
significant}. Although several combining functions are available, mostly due
to its fast convergence to the optimal test when this exists, the one we
consider is $T_{G}=\max (\lambda _{I},\lambda _{S}).$ We observe that since
the rejection region of such a solution is convex in the space of $p$-value
statistics \citep{r22}, then the IU-NPC test $T_{G}$ is a member of a
complete class of test statistics, and so it is \emph{admissible} %
\citep{r40a, r40b}. This means that there does not exist any other combining
function of $(\lambda _{I},\lambda _{S})$ which is uniformly more powerful
than $T_{G}$, unless stringent distributional conditions on data $X$ are
assumed.

\subsection{The permutation solution}

\label{prop}

Suppose, to this end, that $\mathbf{X}_{1}=(X_{11},\ldots,X_{1n_{1}})$ are
the IID \textit{A}-data, and independently $\mathbf{X}_{2}=(X_{21},\ldots
,X_{2n_{1}})\ $the IID \textit{B}-data, so that $\mathbf{X=(X}_{1},\mathbf{X}%
_{2})=(X_{i},i=1,\ldots,n;n_{1},n_{2}),$ where the latter notation means
that first $n_{1}$ elements of pooled set $\mathbf{X}$ are from first sample
and the rest $n_{2}=n-n_{1}$ from the second. So, if $\mathbf{u}^{\ast
}=(u_{1}^{\ast},\ldots,u_{n}^{\ast})$ is any permutation of unit labels $%
\mathbf{u}=(1,\ldots,n)$, the corresponding data permutation is $\mathbf{X}%
^{\ast}=(X(u_{i}^{\ast}),i=1,\ldots,n;n_{1},n_{2}),$ so that $\mathbf{X}%
_{1}^{\ast}=(X(u_{i}^{\ast}),i=1,\ldots,n_{1})$ and $\mathbf{X}%
_{2}^{\ast}=(X(u_{i}^{\ast}),i=n_{1}+1,\ldots,n)$ are the two permuted
samples, respectively. Partial tests $T_{I}^{\ast}=\bar{X}%
_{I2}^{\ast}-X_{I1}^{\ast}$ and $T_{S}^{\ast}=\bar{X}_{S1}^{\ast}-\bar{X}%
_{S2}^{\ast}$ are then calculated on the same permutation of units so as to
obtain their bivariate permutation distribution (steps 6 and 7 of the
algorithm).

The related observed $p$-value statistics are defined as:

\begin{equation*}
\lambda_{h}^{o}=\Pr\{T_{h}^{\ast}(\delta)\geq T_{h}^{o}(\delta)|\mathbf{X}%
_{h}(\delta)\},h=I,S.
\end{equation*}

It is worth noting that such $\lambda _{h}$ were true $p$-values only if the
sharp null $\delta =\delta _{0}$ were\ true. In permutation testing,
however, such quantities are used with the role of statistics that summarize
testing information contained in the observed data $\mathbf{X},$ the most
important property of which are the conditional and unconditional
monotonicity with respect to $\delta .$

It is important to observe that $\lambda_{I}$ and $\lambda_{S},$ being
computed on essentially the same data, are necessarily dependent \citep{r30}%
. Moreover, since they are obtained by means of non-linear transformations
of the data (point 8 of the algorithm), their dependence is generally too
difficult to model and to cope with. It is only known that they are
negatively dependent \citep{r15, r21, r22} and that such a permutational
dependence depends on data $\mathbf{X}$ and margins $(\varepsilon_{I},%
\varepsilon_{S})$. So, unless their bivariate distribution is known,
possibly except for some few estimable nuisance parameters, they must be
combined in a nonparametric way in accordance with the NPC of dependent
tests. Thus, they must be processed simultaneously by means of the same
permutations of units.\medskip\ 

\subsection{An algorithm for the IU-NPC test}

\label{Algorithm} Unless the number of all possible permutations is
relatively small, according to the literature \citep{r8, r10, r42, r21, r22}%
, we estimate the $T_{G}$ distribution by means of a conditional Monte Carlo
procedure, consisting of a random sample of $R$ runs from the set of all
data permutations (commonly, $R$ is set at least equal to $1000$). Two $p$%
-value statistics are then estimated as

\begin{center}
\begin{tabular}{c}
$\hat{\lambda}_{I}^{o}=\sum_{r=1}^{R}\mathbb{I}\mathbf{[}(\bar{X}%
_{I2r}^{\ast }-\bar{X}_{I1r}^{\ast})\geq(\bar{X}_{I2}-\bar{X}_{I1})]/R,$%
\end{tabular}
\end{center}

\noindent and

\begin{center}
\begin{tabular}{c}
$\hat{\lambda}_{S}^{o}=\sum_{r=1}^{R}\mathbb{I}\mathbf{[}(\bar{X}%
_{S1r}^{\ast }-\bar{X}_{S2r}^{\ast})\geq(\bar{X}_{S1}-\bar{X}_{S2})]/R,$%
\end{tabular}
\end{center}

\noindent where $\mathbb{I}$ is the indicator function and $\bar{X}%
_{hjr}^{\ast }=\sum_{i=1}^{n_{j}}X_{hjir}^{\ast }/n_{j},$ $h=I,S,$ $j=1,2$
are calculated at the $r$th permutation, $r=1,\ldots ,R$ (see 
\citet[chapter
6]{r21}; \citet[chapter 4]{r22}). Of course, if the whole permutation space
were inspected, in place of estimations exact numeric values were provided.
\medskip\ 

An algorithm for the IU permutation test is based of the following steps:

\begin{enumerate}
\item read the data set $\mathbf{X=(X}_{1},\mathbf{X}_{2})=(X_{i},i=1,%
\ldots,n;n_{1},n_{2})$ and two margins $\varepsilon_{I}$ and $%
\varepsilon_{S};$

\item define two data vectors $\mathbf{X}_{I}=\mathbf{(X}_{I1},\mathbf{X}%
_{I2})=\mathbf{(}X_{I1i}=X_{1i},i=1,\ldots,n_{1};$ $X_{I2i}=X_{2i}+%
\varepsilon_{I},i=1,\ldots,n_{2})$ and $\mathbf{X}_{S}=\mathbf{(X}_{S1},%
\mathbf{X}_{S2})=\mathbf{(}X_{S1i}=X_{1i},i=1,\ldots,n_{1};$ $%
X_{S2i}=X_{2i}-\varepsilon_{S},i=1,\ldots,n_{2});$

\item compute the observed values of two statistics: $T_{I}^{o}=\bar{X}_{I2}-%
\bar{X}_{I1}$ and $T_{S}^{o}=\bar{X}_{S1}-\bar{X}_{S2}$ and take memory;

\item take a random permutation $\mathbf{u}^{\ast}=(u_{1}^{\ast},\ldots
,u_{n}^{\ast})$ of unit labels $\mathbf{u}=(1,\ldots,n);$

\item define the two permuted data sets: $\mathbf{X}_{I}^{%
\ast}=(X_{I}(u_{i}^{\ast}),i=1,\ldots,n;n_{1},n_{2})$ and $\mathbf{X}%
_{S}^{\ast}=(X_{S}(u_{i}^{\ast}),i=1,\ldots,n;n_{1},n_{2}),$ both defined on
the same permutation $\mathbf{u}^{\ast};$

\item compute the related permuted values of two statistics: $T_{I}^{\ast }=%
\bar{X}_{I2}^{\ast}-\bar{X}_{I1}^{\ast}$ and $T_{S}^{\ast}=\bar{X}%
_{S1}^{\ast}-\bar{X}_{S2}^{\ast}$ and take memory;

\item independently repeat $R$ times steps 4 to 6 obtaining the results: $%
[(T_{Ir}^{\ast},T_{Sr}^{\ast}),$ $r=1,\ldots,R]$ which simulates the
bivariate permutation distribution of two partial tests $(T_{I},T_{S});$

\item calculate two estimates of marginal $p$-value statistics $\hat{\lambda 
}_{I}^{o}=\sum_{r=1}^{R}\mathbb{I}\mathbf{[}T_{Ir}^{\ast}\geq T_{I}^{o}]/R$
and $\hat{\lambda}_{S}^{o}=\sum_{r=1}^{R}\mathbb{I}\mathbf{[}%
T_{Sr}^{\ast}\geq T_{S}^{o}]/R$ and the combined estimated observed value of 
$T_{G}$ as $\hat {T}_{G}^{o}=\max(\hat{\lambda}_{I}^{o},\hat{\lambda}%
_{S}^{o}),$ small values of which are evidence against the null hypothesis $%
H_{0};$

\item if $\hat{T}_{G}^{o}\leq\alpha^{c},$ then reject global $H_{0}$ in
favour of $H_{1},$ i.e. in favour of equivalence.\medskip\ 
\end{enumerate}

It is worth noting that combined test $T_{G}$, in respect to the general NPC
definitions \citep{r22} is nothing else than an \emph{adaptive admissible
combining function}. Then, as such it enjoys all properties of NPC
functions. To be specific, if at least one partial test is consistent, then $%
T_{G}$ is consistent (that property is discussed in Section 2.4). Since
partial tests $T_{I}$ and $T_{S}$ are not positively related, unbiasedness
property must be directly proved. Such a proof simply implies making
reference for two partial tests to the calibrated type I error rates $\alpha
^{c}$ in place of the global $\alpha$ , as discussed in Section 1. It is
also worth noting that rank or other monotonic transformations are to be set
at the end of step 2.

\subsection{A visualization of IU-NPC}

Table \ref{Table1.}, the meaning of symbols being self-evident, provides a
sketch of the IU-NPC procedure, where: $T_{G}^{\ast }$ are obtained
according to an \emph{adaptive weighted rule}, with weights: $w_{h}=1$ if $%
h=\arg \min_{I,S}(T_{I}^{o},T_{S}^{o}),$ and $0$ elsewhere.

So, the observed value of global test is: $%
T_{G}^{o}=w_{I}T_{I}^{o}+w_{S}T_{S}^{o};$ the (empirical) permutation
distribution of which, for $r=1,\ldots,R,$ is: $T_{Gr}^{\ast}=w_{I}T_{Ir}^{%
\ast}+w_{S}T_{Sr}^{\ast}$.

Consequently, the reference $p$-values for $T_{G},$ i.e. $%
\lambda_{G}=\Pr\{T_{G}^{\ast}\geq T_{G}^{o}|\mathbf{X}\}$ are the calibrated
ones $\alpha^{c},$ not $\alpha$. \medskip\ 

\captionof{table}{IU-TOST procedure} \label{Table1.}

\begin{center}
\begin{tabular}{|c||ccccc|}
\hline
$\mathbf{X}$ & $\mathbf{X}_{1}^{\ast }$ & ${\small \cdots }$ & $\mathbf{X}%
_{r}^{\ast }$ & ${\small \cdots }$ & $\mathbf{X}_{R}^{\ast }$ \\ \hline\hline
${\small T}_{I}^{o}$ & ${\small T}_{I1}^{\ast }$ & ${\small \cdots }$ & $%
{\small T}_{Ir}^{\ast }$ & ${\small \cdots }$ & ${\small T}_{IR}^{\ast }$ \\ 
${\small T}_{S}^{o}$ & ${\small T}_{S1}^{\ast }$ & ${\small \cdots }$ & $%
{\small T}_{Sr}^{\ast }$ & ${\small \cdots }$ & ${\small T}_{SR}^{\ast }$ \\ 
\hline
$T_{G}^{o}$ & ${\small T}_{G1}^{\ast }$ & ${\small \cdots }$ & ${\small T}%
_{Gr}^{\ast }$ & ${\small \cdots }$ & ${\small T}_{GR}^{\ast }$ \\ \hline
\end{tabular}
\end{center}

\subsection{Some limiting properties}

Let us assume that population mean $\mathbf{E}_{F}(X)\ $is finite, so that $%
\mathbf{E}(\bar{X}^{\ast}|\mathbf{X})$ is also finite for almost all $%
\mathbf{X}\in\mathcal{X}^{n},$ where $\bar{X}^{\ast}$ is the sample mean of
a without replacement random sample of $n_{1}$ or $n_{2}$ elements from the
pooled set $\mathbf{X},$ taken as a finite population.

Firstly, consider the behavior of partial test $T_{S}^{\ast}(\delta)=\bar {X}%
_{S1}^{\ast}-\bar{X}_{S2}^{\ast},$ where its dependence on effect $\delta$
is emphasized. In \citet{r24}, based on the law of large numbers for
strictly stationary dependent sequences, as are those generated by the
without replacement random sampling (any random permutation is just a
without replacement sample from the pooled data set $\mathbf{X}_{S})$, it is
proved that, as $\min(n_{1},n_{2})\rightarrow\infty,\ $the permutation
distribution of $T_{S}^{\ast}(\delta)$ weakly converges to $\mathbf{E}_{F}(%
\bar{X}_{S1}-\bar{X}_{S2})=(\varepsilon_{S}-\delta)$.

Thus, for any $\delta <\varepsilon _{S}$ the RP of $T_{S}(\delta )$
converges to one: $\mathbf{E}_{F}(\phi _{T_{S}},\delta )\rightarrow 1$.
Moreover, for any $\delta >\varepsilon _{S}$ its RP converges to zero. At
the right extreme of $H_{1S},$ $\delta =\varepsilon _{S}$ say, since for
sufficiently large sample sizes $T_{S}(\varepsilon _{S})$\ rejects with
probability $\alpha ,$ its limit rejection is also $\alpha $.

The behavior of $T_{I}(\delta)$ mirrors that of $T_{S}(\delta)$. That is,
its limiting RP: i) for $\delta=-\varepsilon_{I}$ is $\alpha;$ ii) for $%
\delta<-\varepsilon_{I}$ is zero; iii) for $\delta>-\varepsilon_{I}$ is one.

In the global alternative $H_{1}:(-\varepsilon_{I}<\delta<\varepsilon_{S}),$
since both permutation tests $T_{I}$ and $T_{S}$ are jointly consistent, the
global test $T_{G}$ is consistent too, that is $\mathbf{E}%
_{F}(\phi_{T_{G}},\delta)\rightarrow1$. Correspondingly, for every ($%
\delta<-\varepsilon _{I})\bigcup(\delta>\varepsilon_{S})$ the limiting RP is 
$\mathbf{E}_{F}(\phi_{T_{S}},\delta)\rightarrow0.$ Moreover, in the extreme
points of $H_{0},$ when $\delta$ is either $-\varepsilon_{I}$ or $%
\varepsilon_{S},$ as one and only one can be true if at least one is
positive, the limiting RP of $T_{G}$ is $\alpha$ (if $\varepsilon_{I}=$ $%
\varepsilon_{S}=0,$ this RP is not defined).

\medskip\ 

\begin{figure}[tbp]
\centering
\includegraphics[scale=0.95]{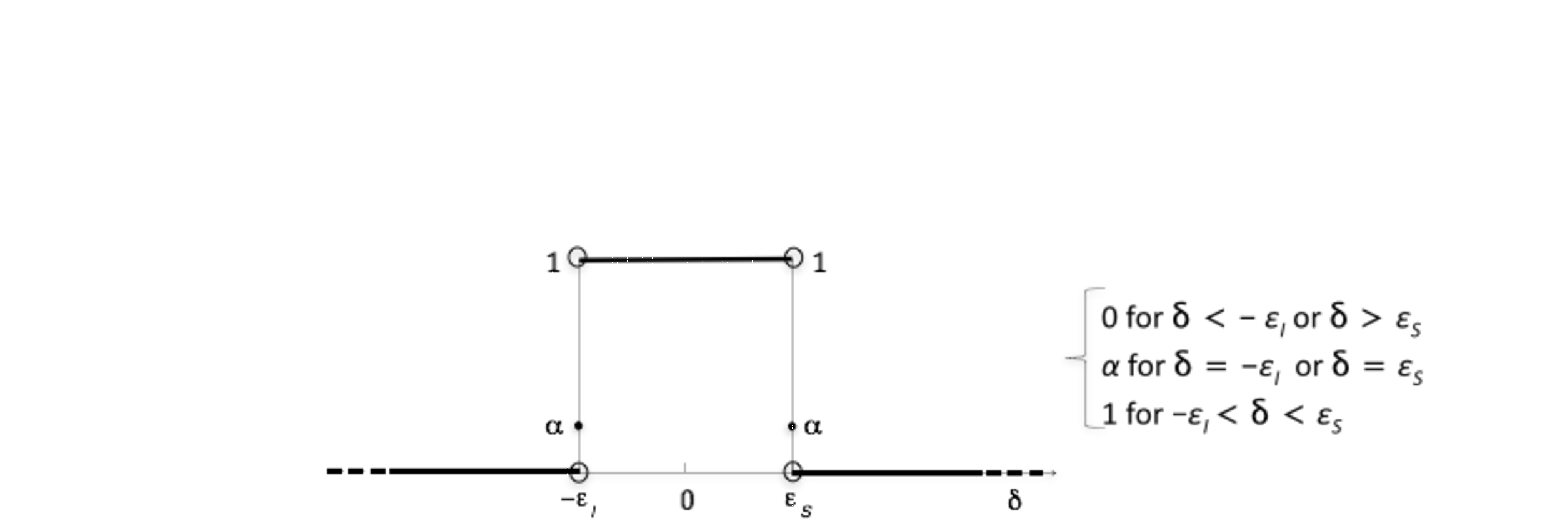} 
\label{Figure 3.}
\caption{Limiting rejection probability of $H_{0}: [\protect\delta \leq -%
\protect\varepsilon_{I}$ OR $\protect\delta \geq \protect\varepsilon_{S}]$ }
\end{figure}

\section{A simple simulation study}

In present section we wish to evaluate the behavior of the IU-NPC
permutation solution both under $H_{0}$ and in power. A comparison with the
optimal likelihood-based competitor $T_{Opt}$ (from \citet{r33}) is also
shown.

Firstly, by using the IU-NPC algorithm with obvious modifications, we report
in Table \ref{Table2.}: the IU-calibrated $\alpha ^{c}$ so as $\alpha
_{G}\approx .05$ for $n_{1}=n_{2}=12,$ $X\sim \mathcal{N}(0,1)$; maximal IU
power $WT_{G}^{\ast }$ at $\delta =0$ and $\alpha _{G}\approx .05, $ maximal
IU power $W\ddot{T}_{G}^{\ast }$ of naive TOST at $\ddot{\alpha}_{I}=\ddot{%
\alpha}_{S}=.05;$ simulations are with $R=5000$ and $MC=10000$.

\newpage $\medskip $ 
\captionof{table}{IU-calibrated $\alpha^{c}$ so as $\alpha_{G}\approx.05$ for $n_{1}=n_{2}=12,$ $X\sim\mathcal{N}(0,1)$; maximal IU power
$WT_{G}^{\ast}$ at $\delta=0$ and $\alpha_{G}\approx.05,$ maximal IU power
$W\ddot{T}_{G}^{\ast}$ of naive IU-TOST at $\ddot{\alpha}_{I}=\ddot{\alpha
}_{S}=.05;$ simulations are with $R=5000$ and $MC=10000$.}\label{Table2.}

\begin{center}
\begin{tabular}{c|ccc}
$\varepsilon _{I}=\varepsilon _{S}$ & $\alpha ^{c}$ & $WT_{G}^{\ast }$ & $W%
\ddot{T}_{G}^{\ast }$ \\ \hline
{\small 0.80~} & {\small 0.060} & {\small 0.301~} & {\small 0.235} \\ 
{\small 0.40~} & {\small 0.185} & {\small 0.076~} & {\small 0.001} \\ 
{\small 0.333} & {\small 0.225} & {\small 0.066~} & {\small 0.000} \\ 
{\small 0.20~} & {\small 0.337} & {\small 0.059~} & {\small 0.000} \\ 
{\small 0.10~} & {\small 0.428} & {\small 0.052~} & {\small 0.000} \\ 
{\small 0.02~} & {\small 0.504} & {\small 0.051~} & {\small 0.000} \\ 
{\small 0.01~} & {\small 0.513} & {\small 0.0505} & {\small 0.000} \\ 
{\small 0.001} & {\small 0.523} & {\small 0.0502} & {\small 0.000}%
\end{tabular}
\end{center}

$\medskip $

\begin{sidewaystable}[p]
\caption{Simulation results for: $X\sim\mathcal{N}(0,1);$
$\alpha=5\%;$ $MC=5000;$ $R=2500;$ $n_{1}=n_{2};$ $\varepsilon_{I}%
=\varepsilon_{S};$ maximal power $W\ddot{T}_{G}^{\ast}$ at $\delta=0$ for
naive IU-TOST $\ddot{T}_{G}^{\ast};$ calibrated partial $\alpha^{c};$ maximal
power $WT_{G}^{\ast}$ for calibrated $T_{G}^{\ast}$; maximal power $WT_{Opt}$
for optimal invariant test $T_{opt}$. }

\medskip
\begin{center}
\label{Table3.}
\begin{tabular}
[c]{|c|ccccc}%

$n$ &  & 10 &  & \multicolumn{1}{c|}{}\\ \hline
$\varepsilon_{I}=\varepsilon_{S}$ & $W\ddot{T}_{G}^{\ast}$ & $\ \alpha^{c}$ &
$WT_{G}^{\ast}$ & $WT_{Opt}$\\\hline
1.0 & ~.392 & .054 & .426 & .453\\
.75 & ~.085 & .078 & .190 & .198\\
.50 & ~.001 & .154 & .091 & .093\\
.25 & ~.000 & .310 & .054 & .058\\
.10 & ~.000 & .434 & .050 & ------
\end{tabular}%
\begin{tabular}
[c]{cccc}
& 15 &  & \multicolumn{1}{c|}{}\\\hline
$W\ddot{T}_{G}^{\ast}$ & $\ \alpha^{c}$ & $WT_{G}^{\ast}$ & $WT_{Opt}$\\\hline
~.704 & .050 & .704 & .714\\
~.040 & .059 & .348 & .355\\
~.008 & .113 & .123 & .127\\
~.000 & .271 & .061 & .063\\
~.000 & .417 & .053 & ------
\end{tabular}%
\begin{tabular}
[c]{cccc}
& 20 &  & \\\hline
$W\ddot{T}_{G}^{\ast}$ & $\alpha^{c}$ & $WT_{G}^{\ast}$ & $WT_{Opt}$\\\hline
~.846 & .050 & .846 & .859\\
~.513 & .052 & .527 & .533\\
~.032 & .084 & .163 & .171\\
~.000 & .237 & .065 & .068\\
~.000 & .402 & .056 & ------
\end{tabular}
\end{center}
\end{sidewaystable}

These results confirm that calibrated $\alpha ^{c}$ lie in the half-open
interval $[\alpha ,~(1+\alpha )/2),$ and that for margins $\mathbf{\
\varepsilon }$ smaller than about $\sigma _{X}/3 $ the maximal power of
naive TOST $\ddot{T}_{G}$ is close to zero (in first three decimal figures;
moreover, there are situations where it can be exactly zero). The latter
justify our sentence that the naive TOST with moderate sample sizes and
margins never can find that a drug is equivalent to itself; i.e. the spirit
of point V in Section 1 is largely confirmed. In particular, for naive $%
\ddot{T}_{G}$ to be unbiased, i.e. for its power is at least $.05$,\ when $%
\varepsilon _{I}=\varepsilon _{S}=0.2$ sample sizes of $n_{1}=n_{2}\approx
280$ are needed. In our opinion, these facts suggest to abandon the naive
TOST solution in the analysis of practical equivalence problems.

Table \ref{Table3.} report simulation results for: $X\sim \mathcal{N}(0,1);$ 
$\alpha =5\%;$ $MC=5000;$ $R=2500;$ $n_{1}=n_{2};$ $\varepsilon
_{I}=\varepsilon _{S};$ maximal power $W\ddot{T}_{G}^{\ast }$ at $\delta =0$
for naive TOST $\ddot{T}_{G}^{\ast };$ calibrated partial $\alpha ^{c};$
maximal power $WT_{G}^{\ast }$ for calibrated $T_{G}^{\ast }$; maximal power 
$WT_{Opt}$ for optimal invariant test $T_{Opt}$ (the latter are from 
\citet[page
122]{r33})$.$

Latter results confirm optimality of $T_{Opt}$. However, performances of $%
T_{Opt}$ and of IU-NPC $T_{G}$ are comparable and their power are quickly
converging according to increasing sample sizes (point iv in Section 1).
Also confirmed is that power of naive $\ddot{T}_{G}$ converges to that of
calibrated IU-NPC $T_{G}$ as margins increase and that both tend to one
according to Berger's Theorem 2 \citep{r1}.

The same IU-NPC simulation algorithm can also be used for determining the
design $n_{1}=n_{2}$ such that Max $WT_{G}=p$ at standardized margins $%
\varepsilon _{I}=\varepsilon _{S} $ and calibrated $\alpha ^{c}=\alpha $.
The following table contains some few designs obtained by assuming: $X\sim 
\mathcal{N}(0,1);$ $\alpha =5\%;$ $p=0.80;$ $MC=5000;$ $R=2500.$

\begin{table}[!h]
\caption{Designs obtained by assuming: $X\sim \mathcal{N}(0,1);$ $\protect%
\alpha =5\%;$ $p=0.80$.}
\label{Table_Nuova1.}
\medskip \centering
\begin{tabular}{c|cccccc}
${\small \varepsilon }_{I}{\small =\varepsilon }_{S}$ & {\small 1.00} & 
{\small 0.80} & {\small 0.60} & {\small 0.40} & {\small 0.20} & {\small 0.10}
\\ 
&  &  &  &  &  &  \\ 
${\small n}_{1}{\small =n}_{2}$ & {\small 18} & {\small 28} & {\small 49} & 
{\small 109} & {\small 435} & {\small 1738}%
\end{tabular}%
\end{table}

Assuming that at $\varepsilon =1$ the sample size, as obtained by
interpolating the entire simulation results, is $n(1)=17.38,$ designs for $%
\varepsilon =(0.40,0.20,0.10)$ have been obtained according to the rule $%
n(\varepsilon ^{\prime })=(1/\varepsilon ^{\prime })^{2}n(1).$ This same
rule can also be used for deducing\ all intermediate designs$.$ It is worth
noting that these designs are strictly close to those obtained within the
naive TOST approach as reported in Lakens (2017). Such a practical
coincidence is mostly due to the fact that calibrated $\alpha ^{c}$
coincides with non-calibrated $\alpha $ for interval length, adjusted with
sample sizes, of about $(\varepsilon _{I}+\varepsilon _{S})\sqrt{%
n_{1}n_{2}/n\sigma ^{2}}>5.4.$

\section{Three application examples}

With the role of putting into evidence that IU-NPC requires quite large
margins to detect equivalence, we report the analyses of three examples. The
data of the first do manifest a clear equivalence of two distributions since
their sample means lie within a reasonably small interval, those of the
second manifest a practical equivalence, instead those of the third clearly
manifest a substantial non-equivalence.

\medskip\ 

Example 1. On sulphur content in two batches of raw material ($n_{1} =n_{2}
=20)$ , from Anderson-Cook \& Borror (2016). The data are:\medskip\ 

\begin{center}
\begin{tabular}{ccccccccc}
& I &  &  &  &  &  & II &  \\ \cline{1-3}\cline{7-9}
{\small 0.4889} &  & {\small 0.5214} &  &  &  & {\small 0.4823} &  & {\small %
0.5073} \\ 
{\small 0.4818} &  & {\small 0.5031} &  &  &  & {\small 0.5165} &  & {\small %
0.5154} \\ 
{\small 0.5123} &  & {\small 0.4451} &  &  &  & {\small 0.4622} &  & {\small %
0.4671} \\ 
{\small 0.4688} &  & {\small 0.4951} &  &  &  & {\small 0.4853} &  & {\small %
0.5426} \\ 
{\small 0.4575} &  & {\small 0.4684} &  &  &  & {\small 0.4768} &  & {\small %
0.5272} \\ 
{\small 0.5238} &  & {\small 0.4853} &  &  &  & {\small 0.4984} &  & {\small %
0.4889} \\ 
{\small 0.4483} &  & {\small 0.4558} &  &  &  & {\small 0.5224} &  & {\small %
0.4871} \\ 
{\small 0.5346} &  & {\small 0.4842} &  &  &  & {\small 0.4889} &  & {\small %
0.4872} \\ 
{\small 0.4851} &  & {\small 0.4726} &  &  &  & {\small 0.4564} &  & {\small %
0.4920} \\ 
{\small 0.4818} &  & {\small 0.5257} &  &  &  & {\small 0.5028} &  & {\small %
0.5291}%
\end{tabular}
\end{center}

\medskip\ 

It is asked to establish if sulfur content is equivalent on two batches.

Basic statistics are: $\bar{X}_{1}=0.487$, $\bar{X}_{2}=0.497$, $\hat{\sigma}%
_{1}=0.0265$, $\hat{\sigma}_{2}=0.0234$, $\hat{\sigma}=0.0252$.

For two-sided (sharp) hypotheses $H_{0}^{\prime }:X_{1}\overset{d}{=}X_{2}$
V.s $H_{1}^{\prime }:X_{1}\overset{d}{\neq }X_{2},$with $R=100000$, PT $T=|%
\bar{X}_{1}-\bar{X}_{2}|$ \ the $p$-value statistic is $\hat{\lambda}%
=0.2221; $ a value that manifest a substantial equivalence (Eq) of two
distributions.

The results of our IU-NPC analysis for margins $\varepsilon _{I}=\varepsilon
_{S}=(0.005$, $0.010$, $0.020$, $0.0232$, $0.0239$, $0.025),$ corresponding
to standardized values (in terms of $\hat{\sigma})$ of ($0.198,$ $0.397,$ $%
0.794,$ $0.921,$ $0.950,$ $0.992$), for respectively original data $\mathbf{X%
}$ and their mid-ranks $\mathbf{MR}$ are reported in the following
table:\medskip\ 

\begin{center}
\begin{tabular}{llllll}
&  &  &  & ~$\mathbf{X}$~ &  \\ 
$\varepsilon _{I}=\varepsilon _{S}$ & {\small \ }$\alpha ^{c}$ &  & {\small %
\ }$\hat{\lambda}_{G}$ &  & \ {\small Inference} \\ \cline{1-2}\cline{4-6}
{\small 0.005} & {\small 0.301} &  & {\small 0.727} &  & $H_{0}:${\small \ \
N-Eq} \\ 
{\small 0.010} & {\small 0.126} &  & {\small 0.491} &  & $H_{0}:${\small \ \
N-Eq} \\ 
{\small 0.020}$^{(\dag )}$ & {\small 0.052} &  & {\small 0.103} &  & $H_{0}:$%
{\small \ \ N-Eq} \\ 
{\small 0.0232} & {\small 0.050} &  & {\small 0.0494} &  & $H_{1}:${\small \
\ Eq} \\ 
{\small 0.0239} & {\small 0.050} &  & {\small 0.0421} &  & $H_{1}:${\small \
\ Eq} \\ 
{\small 0.025} & {\small 0.050} &  & {\small 0.031} &  & $H_{1}:${\small \ \
Eq}%
\end{tabular}%
\qquad 
\begin{tabular}{lll}
& $\mathbf{MR}$ &  \\ 
$\hat{\lambda}_{RG}$ &  & {\small \ \ Inference} \\ \hline
{\small 0.698} &  & $H_{0}:${\small \ \ N-Eq} \\ 
{\small 0.461} &  & $H_{0}:${\small \ \ N-Eq} \\ 
{\small 0.113} &  & $H_{0}:${\small \ \ N-Eq} \\ 
{\small 0.055} &  & $H_{0}:${\small \ \ N-Eq} \\ 
{\small 0.050} &  & $H_{1}:${\small \ \ Eq} \\ 
{\small 0.045} &  & $H_{1}:$ {\small \ Eq}%
\end{tabular}%
\medskip\ 
\end{center}

Note: $\varepsilon _{I}=\varepsilon _{S}=0.02^{(\dag )}$ (corresponding to a
standardized value of $0.794$) are the margins adopted by Anderson-Cook \&
Borror in their analyses while adopting the naive TOST test $\ddot{T}_{G}$
based on Student's $t$ distribution. It is worth observing that their
corresponding non-calibrated \emph{p}-value is of $0.103$, the same as
calibrated ours based on permutations.

Permutation \emph{p}-value statistics were obtained with $R=100000$ random
permutations.

Calibrated $\alpha ^{c}$, corresponding to $\alpha _{G}=0.05,$ that cannot
be exactly determined since the underlying distribution $F$ is not known
(point ii in Section 1), were assessed assuming validity of the permutation
central limit theorem, leading to approximate partial test distributions by
assuming normal laws for the data, i.e. $Y_{h}\sim $ $\mathcal{N}(\mu
_{h}=\varepsilon _{h}{/\hat{\sigma}},~\sigma _{h}=\hat{\sigma}),$ ${%
h=-\varepsilon }_{I},\varepsilon _{S}$, with $5000$ Monte Carlo simulations
and $R=2500$ random permutations each.

Mid-ranks are used in place of plain ranks to reduce the impact of ties;
indeed, when there are no ties mid-ranks and plain ranks give exactly the
same results.

The IU-NPC on original data $\mathbf{X}$ accepts non-equivalence (N-Eq) for
all standardized margins $\varepsilon <0.921$ and equivalence (Eq) for $%
\varepsilon >0.921.$Of course, mid-rank based results reflect the same
behavior as those on original data except that, to obtain corresponding
inferences, slightly larger margins and/or sample-sizes are apparently
required. In particular it is worth observing that non-equivalence N-Eq is
obtained for margins up to $0.950$.

In our opinion, standardized margins of $0.921$ for original data $\mathbf{X}
$ and of $0.950$ for mid-ranks are too large for meaningful practical
applications of equivalence testing in the area of quality control.

These IU-NPC results, however, manifest severe difficulties for $T_{G}$ to
detect a substantial equivalence when it is really evident in practice.

\medskip\ 

\textbf{Example 2.} Consider the data from \citet{r41} on the end-point
variable $Log\ C_{\max }$, related to $n_{1}=20$ Japanese subjects and $%
n_{2}=13$ Caucasians, after prescribing a drug. Data concern a bridging
study conducted to investigate for bio-equivalence between two populations.
So, it is asked to test if two populations can be retained as bio-equivalent
with respect to that variable. Data are in Table \ref{Table4.}.

The basic statistics with these data are: $\bar{X}_{Jap}=1.518;$ $\hat{%
\sigma }_{Jap}=0.0812;$ $\bar{X}_{Cau}=1.457;$ $\hat{\sigma}_{Cau}=0.0951;$
pooled $\hat{\sigma}=0,0869.$

By firstly using the permutation test $T^{\ast }=|\bar{X}_{J}^{\ast }-\bar{X}%
_{C}^{\ast }|$ for the sharp null hypothesis\ $H_{0}^{\prime }:X_{J}\overset{%
d}{=}X_{C}$ against the two-sided alternative $H_{1}^{\prime }:X_{J}\overset{%
d}{\neq }X_{C}$, with $R=100000$ we obtain the $p$-value statistic $\hat{%
\lambda}=0.0535$ (for the one-sided $H_{1}^{\prime \prime }:X_{J}\overset{d}{%
>}X_{C}$ it is $\hat{\lambda}=0.0268$; with Fisher-Mood's median test the
one-sided exact $p$-value is $\lambda =0.0581)$. Thus, denoting a practical
equivalence between two data sets at $\alpha =5\%$, although $\bar{X}_{Jap}$
appears to be slightly larger than $\bar{X}_{Cau}.$

Let us consider the IU-NPC $T_{G}$ for testing equivalence with a list of
margins $\varepsilon _{I}=\varepsilon _{S}=(0.022,$ $0.058,$ $0.071,$ $%
0.109, $ $0.120,$ $0.125),$ approximately corresponding to $(1/4,$\ $2/3,$ $%
0.82,\ 1.25,$ $1.38,$ $1.44)$ times the pooled $\hat{\sigma}=0.0869$,
respectively.\medskip\ 

\begin{center}
\captionof{table}{Data  of Example 2 \cite{r41}.}\label{Table4.} 
\begin{tabular}{ccccccccccc}
& \multicolumn{1}{|c}{\small 1.567} & {\small 1.515} & {\small 1.500} & 
{\small 1.591} & {\small 1.624} & {\small 1.691} & {\small 1.531} & {\small %
1.456} & {\small 1.351} & {\small 1.478} \\ 
{\small Jap} & \multicolumn{1}{|c}{\small 1.461} & {\small 1.571} & {\small %
1.565} & {\small 1.586} & {\small 1.406} & {\small 1.488} & {\small 1.500} & 
{\small 1.577} & {\small 1.500} & {\small 1.407} \\ 
&  &  &  &  &  &  &  &  &  &  \\ 
{\small Cau} & \multicolumn{1}{|c}{\small 1.455} & {\small 1.375} & {\small %
1.474} & {\small 1.650} & {\small 1.464} & {\small 1.375} & {\small 1.479} & 
{\small 1.413} & {\small 1.423} & {\small 1.389} \\ 
& \multicolumn{1}{|c}{\small 1.441} & {\small 1.650} & {\small 1.348} &  & 
&  &  &  &  & 
\end{tabular}%
\medskip\ 
\end{center}

The results, with $R=100000$ on original data $X$ and their mid-rank
transformations $MR$ respectively are:\medskip\ 

\begin{center}
\begin{tabular}{llllll}
&  &  &  & ~$\mathbf{X}$~ &  \\ 
$\varepsilon _{I}=\varepsilon _{S}$ & $\ \alpha ^{c}$ &  & {\small \ }$\hat{%
\lambda}_{G}$ &  & {\small \ \ Inference} \\ \cline{1-2}\cline{4-6}
{\small 0.022} & {\small 0.264} &  & {\small 0.902} &  & $H_{0}:${\small \ \
N-Eq} \\ 
{\small 0.058} & {\small 0.068} &  & {\small 0.545} &  & $H_{0}:${\small \ \
N-Eq} \\ 
{\small 0.071} & {\small 0.050} &  & {\small 0.382} &  & $H_{0}:${\small \ \
N-.Eq} \\ 
{\small 0.109} & {\small 0.050} &  & {\small 0.071} &  & $H_{1}:${\small \ \
N-Eq} \\ 
{\small 0.120} & {\small 0.050} &  & {\small 0.039} &  & $H_{1}:${\small \ \
Eq} \\ 
{\small 0.125} & {\small 0.050} &  & {\small 0.025} &  & $H_{1}:${\small \ \
Eq}%
\end{tabular}%
\qquad 
\begin{tabular}{lll}
& $\mathbf{MR}$ &  \\ 
$\hat{\lambda}_{RG}$ &  & {\small \ \ Inference} \\ \hline
{\small 0.960} &  & $H_{0}:${\small \ \ N-Eq} \\ 
{\small 0.720} &  & $H_{0}:${\small \ \ N-Eq} \\ 
{\small 0.600} &  & $H_{0}:${\small \ \ N-.Eq} \\ 
{\small 0.154} &  & $H_{0}:${\small \ \ N-Eq} \\ 
{\small 0.063} &  & $H_{0}:${\small \ \ N-Eq} \\ 
{\small 0.039} &  & $H_{1}:${\small \ \ Eq}%
\end{tabular}%
\medskip\ 
\end{center}

At $\varepsilon $ such that $\ddot{\alpha}(\varepsilon )=\alpha _{G}=0.05$,
i.e. $\varepsilon \approx 0.071$ (corresponding to $\approx 0.82~\hat{\sigma}%
),$ type I error rates of naif $\ddot{T}_{G}(\varepsilon )$ and of $T_{G}$
approximately coincide, since $\alpha ^{c}\approx \ddot{\alpha}\approx
\alpha $. Of course, this coincidence remains also for larger margins and
sample sizes. With the data of the example, the equivalence\ of two data
sets is accepted if margins $\varepsilon _{I}=\varepsilon _{S}\gtrsim 1.38~%
\hat{\sigma}$. In our opinion, these too wide margins might be considered as
an extremely poor result which puts into evidence a known characteristic
difficulty of the IU-TOST approach, as well as that of\ the likelihood-based
one, while detecting for equivalence especially when it practically is.

\medskip\ 

\textbf{Example 3.} Data, from \citet{r22}, are related to a psychological
experiment on job satisfaction of $n=20$ workers in a company, where $%
n_{1}=12$ were classified as Extroverted and $n_{2}=8$ as Introverted. Some
criteria for equivalence analysis of psychological data can be found, for
instance, in \cite{r3n}. Data are in Table \ref{Table5.} Basic statistics
are: $\bar{X}_{1}=65.92;$ $\hat{\sigma}_{1}=8.61;$ $\bar{X}_{2}=48.63;$ $%
\hat{\sigma}_{2}=9.44;$ pooled $\hat{\sigma}=8.93.$ For the two-sided
(sharp) hypotheses $H_{0}:X_{1}\overset{d}{=}X_{2}$ \ V.s \ $H_{1}:X_{1}%
\overset{d}{\neq }X_{2},$ with $R=100\,000$, PT $T=|\bar{X}_{1}-\bar{X}_{2}|$
leads to $\hat{\lambda}=0.00086$ which manifests a substantial
non-equivalence. The IU-NPC for equivalence with $R=100000,$ using $%
\varepsilon _{I}=\varepsilon _{S},$ gives results in Table \ref{Table6.}
Since for margins $\varepsilon \geq 15,$ calibrated $\alpha ^{c}$ is
approximately equal to $\alpha =0.05,$ the IU-NPC results permit declaring
non-equivalence for margins $\varepsilon \leq 24$ and equivalence for larger
values, when data $X$ lie in the range $\bar{X}\pm 2.7~\hat{\sigma},$ i.e. $%
\approx 59\pm 25$.\medskip\ 

\captionof{table}{Data of Example 2 \cite{r22}.}\label{Table5.} 
\begin{tabular}{|lll}
{\small Extroverted:} & $\mathbf{X}_{1}=$ {\small (66, 57, 81, 62, 61, 60,
73, 59, 80, 55, 67, 70)} & ${\small n}_{1}{\small =12}$ \\ 
{\small Introverted:} & $\mathbf{X}_{2}=$ {\small (64, 58, 45, 43, 37, 56,
44, 42)} & ${\small n}_{2}{\small =8}$%
\end{tabular}%

\bigskip
Mid-rank-based results in Table \ref{Table6.} reflect those on original
data. Of course, when standardized mean difference is large, ($%
65.92-48.63)/8.93\approx 1.94,$ say), IU-NPC detects non-equivalence with a
probability larger than $\alpha $ (the approximate maximal power of N-Eq in
this framework would be of about $0.982$).

However, since from in \cite{r3n} for psychological experiments the
suggested equivalence margins are of $\varepsilon \approx 0.1\cdot $\/$%
\sigma \approx 0.9,$ which would correspond a maximal power of about $0.052$
with calibrated $\alpha ^{c}=0.437.$ Indeed, a very poor performance in that
discipline. Moreover and considerably important in our opinion, on three
examples once $H_{0}:$ N-Eq has been accepted it is unclear how to proceed
for making inference on which of two arms, $H_{0I}$ or $H_{0S}$, is active
while controlling type I errors especially when calibrated $\alpha ^{c}\ $is
larger than nominal $\alpha $.\medskip\ 

\captionof{table}{IU-NPC for equivalence with $R=100000,$ using
$\varepsilon_{I}=\varepsilon_{S}$}\label{Table6.}

\begin{center}
\begin{tabular}{cccccc}
&  &  &  & ~$\mathbf{X}$~ &  \\ 
${\small \varepsilon }_{I}{\small =\varepsilon }_{S}$ & $\alpha ^{c}$ &  & $%
\hat{\lambda}_{G}$ &  & {\small Inference} \\ \cline{1-2}\cline{4-6}
{\small 22} & {\small 0.05} &  & {\small 0.136} &  & $H_{0}:$ \ {\small N-Eq}
\\ 
{\small 24} & {\small 0.05} &  & {\small 0.062} &  & $H_{0}:$ \ {\small N-Eq}
\\ 
{\small 25} & {\small 0.05} &  & {\small 0.035} &  & $H_{1}:$ \ {\small Eq \
\ \ }%
\end{tabular}%
\qquad 
\begin{tabular}{ccc}
& $\mathbf{MR}$ &  \\ 
$\hat{\lambda}_{RG}$ &  & {\small Inference} \\ \hline
{\small 0.164} &  & $H_{0}:$ \ {\small N-Eq} \\ 
{\small 0.054} &  & $H_{0}:$ \ {\small N-Eq} \\ 
{\small 0.026} &  & $H_{1}:$ \ {\small Eq \ \ \ }%
\end{tabular}%
\medskip\ 
\end{center}

\section{Conclusions}

From all three examples it results that IU approaches (likelihood-based,
naive TOST, and NPC calibrated) apparently try to preserve the
non-equivalence conclusion even when this is evidently not true. Moreover,
from simulations it results that\ their power in detecting equivalence when
it is really true is generally too poor, thus implying quite severe
inferential costs.

The nonparametric combination (NPC) of dependent permutation tests, when the
permutation testing principle applies, enables us dealing with the rather
intriguing problem of testing for equivalence and non-inferiority in a
general unidimensional setting according to the IU-NPC approach. Two related
crucial points, as pointed out by \citet{r30}, are how to go beyond\ the
likelihood ratio methods, which are generally too difficult to apply
properly, and how to do with the generally too complex dependence structure
of the two partial test statistics in which such an analysis is usually
broken down. Using the results and methods discussed in the books of %
\citet{r21} and \citet{r22} concerning the NPC we are able to provide a
general solution to the testing under the TOST approach which rationally can
interpret one of the ways to face the equivalence and non-inferiority
problem.

Extensions to multivariate settings (see for example \citet{r1r}),
especially when, for some of the variables, equivalence effects are in
hyper-rectangular margins and others are hyper-unidirectional (for instance,
as with side effects of drugs for which it is required to be not larger than
a target) as well as extensions to one sample designs, to $C>2$ samples, to
ordered categorical endpoint variables, to repeated measurements, and to
some situations where missing and/or censored data are informative on
treatment effects, will be the subject matters of future researches. We
expect that these extensions can be obtained by suitable adaptive
modifications of the combining functions with respect to the corresponding
solutions discussed in \citet{r21, r23} and \citet{r2r} on \emph{%
multi-aspect testing} and the NPC \citep{r43}. Another promising research
field where the method we proposed could be effectively applied, is that one
of statistical process control; in particular, the IU permutation solution
for two-sample equivalence testing could be helpful to face the problem of
ranking of several industrial product/prototypes \citep{r3r} and monitoring
industrial processes in case of multivariate responses \citep{r4r}.

Since the equivalence problem at hand can find appropriate solution also
within the Union-Intersection (UI) approach, as is done in \cite{r25}, we
postpone to a further paper a parallel analysis of IU-NPC and UI-NPC
permutation solutions.

\section*{Acknowledgements}

The authors express their thanks to two anonymous referees and the
associated editor for their valuable comments and criticism that contribute
to several improvements of our paper. This research received no specific
grant from any funding agency in the public, commercial, or not-for-profit
sectors. All authors have equally contributed to the research.

\end{document}